# OSCILLOMETRIC CONTINUOUS BLOOD PRESSURE SENSING FOR WEARABLE HEALTH MONITORING SYSTEM

Gennaro Gelao[1], Roberto Marani[1], Vittorio M. N. Passaro[2] and Anna Gina Perri[1]

[1]Electronic Devices Laboratory, [2]Photonics Research Group,
[1,2] Electrical and Information Engineering Department, Polytechnic University of Bari,
via E. Orabona 4, Bari – Italy
annagina.perri@poliba.it

## ABSTRACT

*In this paper we present an acquisition chain for the measurement of blood arterial pressure based on the oscillometric method. This method does not suffer from any limitation as the well-known auscultatory method and it is suited for wearable health monitoring systems. The device uses a pressure sensor whose signal is filtered, digitalized and analyzed by a microcontroller. Local analysis allows the evaluation of the systolic and diastolic pressure values which can be used for local alarms, data collection and remote monitoring.*

## 1. INTRODUCTION

The most recent developments of in the fields of Electronics, Informatics and Telecommunications include applications in the biomedical engineering field to improve the healthcare quality [1-8]. In particular, a number of systems has been developed in telemedicine and home care sectors which could guarantee an efficient and reliable home assistance, allowing a highly better quality of life in terms of prophylaxis, treatment and reduction of discomfort connected to periodic hospitalization for the patients afflicted by pathologies, such as hypertension, as well as considerable savings on sanitary expenses. In particular, hypertension is defined as elevated blood pressure (BP) above 140 mm Hg systolic and 90 mm Hg diastolic when measured under standardized conditions [9-10]. Hypertension can be a separate chronic medical condition estimated to be affecting a quarter of the world's adult population [11], as well as a risk factor for other chronic and non-chronic patients. Traditional high-risk patients include all patients afflicted by pathologies such as cardiac decompensation, ischemic heart disease, kidney disease, diabetes. Persistent hypertension is one of the key risk factors for strokes, heart attacks and increased mortality [12]. In particular, in pregnant women with gestational diabetes, known as preeclampsia, hypertension is the most common cause of maternal and fetal death [13]. For all the previous cases blood pressure should be kept below 130 mmHg systolic and 80 mm Hg diastolic to protect the kidneys from BP-induced damage [14].

Therefore, in particular for high-risk patients, it is very important to continuously monitor the blood pressure over a whole day using systems that should not hamper the ordinary daily activities of the patient. By the data collected during continuous monitoring, physician can have a clear understanding of the daily evolution of the pressure values, furthermore these devices can trigger out-of-range alarms to alert both the patient and the physician.

Continuous monitoring is also essential to rule out so called white-coat hypertension where an healthy patient seems having hypertension just because of the stress during the physician visit.

The most used method for measuring blood pressure is the auscultatory method. This method is based on the contemporary use of a sphygmomanometer and a stethoscope. The sphygmomanometer has a cuff, which inflates and deflates, equipped with a pressure sensor positioned on the arm in correspondence of the brachial artery. The stethoscope allows to listen the arterial sounds (known as Korotkoff sounds) during the cuff slow deflation, which are used to determine systolic and diastolic blood pressure. This method presents some difficulties in signal analysis due to physiological variations of the Korotkoff sound patterns. Moreover, weak signals are disturbed by ambient noises and misleading information can occur [15]. This method gives very few data samples of the BP and furthermore requires the presence of qualified operator, while the daily activities of the patient are suspended for a time longer than the same measurement. The recalled technical difficulties make the method automation quite hard to be achieved when good data quality is required. Instead of the auscultatory method, there are other important indirect methods such as the oscillometric method, which is one of the best approach to evaluate the systolic and diastolic blood pressure [16].

In this paper we propose an electronic unit to perform a non-invasive measurement of the blood pressure based on the oscillometric method and able to evaluate both the systolic and diastolic blood pressure values. To avoid artifacts, accelerometers could be used to verify that the patient is in rest and that the arm is lowered.

The proposed system, prototyped and tested at the Electronic Devices Laboratory, is characterized by easy use and very high level of automation. The paper is organized as follows.

In Section 2 we describe shortly the main features of our system, while in Section 3 the proposed circuit is analyzed, highlighting the main goals obtained by our design. Finally, the conclusions and future developments are illustrated in Section 4.

## 2. MAIN FEATURES OF THE PROPOSED SYSTEM

As we have already considered, the proposed system is based on the oscillometric method [16]. This approach analyzes the variations in pulse pressure as a function of the pressure applied to a pneumatic cuff wrapped around the limb. As in the auscultatory method, the cuff is inflated until the artery is completely occluded. A stepwise decrease in cuff pressure is then applied, and an increase in pulse amplitude is observed when the cuff pressure equals the blood systolic pressure. The pulse amplitude increases until the mean blood pressure is reached. The pulse amplitude then decreases with decreasing of the cuff pressure from mean to diastolic values. The systolic and diastolic blood pressure can then be evaluated by applying a numerical algorithm to the shape of oscillometric amplitudes. Moreover, this method allows the measurement of the blood pressure also when the Korotkoff sounds are weak, thus overcoming the limitations related to the auscultatory method. Since microphones are not required, environmental noise problems are completely avoided.

The acquisition chain consists of a pressure transducer whose signal is amplified and filtered to cut noise and continuous levels, then the signal is read by an analog to digital converter (ADC) on a microcontroller. The microcontroller, after deflation of the cuff, allows to determine the peak of the pulsatile component and the diastolic and systolic pressure values. Finally, the microcontroller runs the memory and wireless transmitter/receiver unit, including the battery.

In particular, the combination of the latest suitable telecommunication solutions (GPRS and Bluetooth) with new algorithms and solutions for automatic real-time, cost-effective diagnosis (both in terms of purchase expenses and data transmission/analysis) and simplicity of use (the patient will be able to wear it), can make the designed system useful for remote monitoring, allowing real-time rescue operations in case of emergency without the necessity for data to be constantly monitored.

To this purpose, the proposed system has been equipped with properly developed firmware [1-2], [17], which enables automated functioning and complex decision-making. It is indeed able to prevent lethal risks thanks to an automatic warning system. Everything occurs automatically, without any operation by the user.

Each monitored patient is given a case sheet on a Personal Computer (PC) functioning as a server (online doctor). Data can also be downloaded by any other PC, palmtop or smart phone equipped with a browser. The system reliability rests on the use of a distributed server environment, which allows its functions not to depend on a single PC and gives more online doctors the chance to use them simultaneously. The system consists of three hardware units as well as properly developed management software, including:

- the cuff wrapped at an arm;
- a wearable Portable Unit (PU), which is wireless (GPRS/Bluetooth). This PU allows, by an Internet connection, the transmission, continuous or sampled or on demand, of the health parameters and allows the GPS satellite localization and the automatic alarm service, on board memory. Moreover, PU has an USB port for data transfer and a rechargeable battery;
- Relocable Unit (RU): GPRS/Bluetooth Dongle (on PC server, i.e. online doctor):
- Management Software: GPS mapping, address and telephone number of nearest hospital, simultaneous monitoring of more than one patient, remote (computerized) medical visits and consultation service, creation and direct access to electronic case sheets (login and password).

## 3. ANALYSIS OF THE PROPOSED CIRCUIT

As a first step of the measurements, the cuff wrapped around the patient arm is inflated with air by a micro-pump and then deflated through an electromechanical valve. Inside the cuff, facing the arm, a sensor measures the pressure obtained with summing to the arterial pressure variation (oscillometric signal) produced by the patient's heartbeat.

During deflation, a microcontroller senses the pressure variation and identifies the instants when systolic and diastolic blood pressure are reached, so at those times the pressure is read from the pressure sensor and the measurement is achieved. The system includes several subunits, but we will focus mainly on the acquisition chain.

A transducer is used to sense the pressure inside the cuff. For higher sensitivity we have chosen a piezoresistive micromachined transducer with analog output. It is an integrated transducer constituted by two gain stages, where the first provides the temperature compensation. The sensor is characterized by an input pressure range [0 ÷ 50] kPa, an output voltage range [0.2 ÷ 4.7] V, a sensitivity of 90 mV/kPa = 12 mV/mmHg, a supply voltage of 5 V and supply current of 7 mA. Since the medium arterial pressure is in the range [1 ÷ 3] mmHg, we obtain an output voltage range [12 ÷ 36] mV. We have chosen to have large margins in electronic design for hearth rate and pressure signal, so that the system can be applied to young people under physical stress as to unhealthy people with high pressure values. In Fig. 1 the signal coming from the piezoresistive transducer is shown, where two curves can be distinguished, corresponding to inflating and deflating processes, respectively. We have analyzed the signal during the deflating process, which occurs after the first six seconds. The application of the Fourier transform shows the signal frequency components, as sketched in Fig. 2.

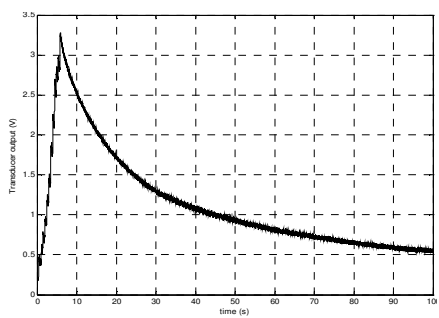
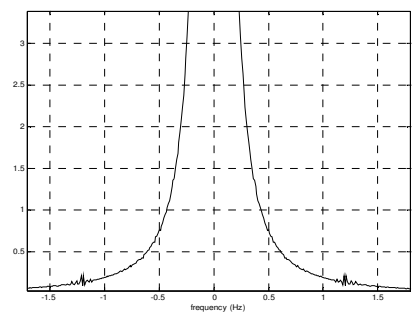

Figure 1. Pressure signal behavior.          Figure 2. Pressure signal spectral components.

Thus, this signal is the superposition of a deflation signal with the arterial pressure pulsation (oscillometric) signal. While the former is a very low frequency signal, the latter has base frequency at the earth beat [1 ÷ 3] Hz, so a pass-band filter with a low cut-off frequency is used to separate the oscillometric signal. The pressure is obtained by the complete signal, where the small oscillating component of the oscillometric signal is easily suppressed by averaging.

Hence the signal from the transducer is split in two paths, one path reaches straight the ADC of the microprocessor, while a second path passes through a filter before arriving to a second ADC of the microprocessor. The former transmits the pressure values, while the latter transmits the pressure variations due to the earth beat.

Now, on the filtered path we have to eliminate the DC component, out-band noise and interferences from other electrical devices at higher frequencies. To this aim, a first order high pass filter with a gain of 24 is used as first stage, followed by a second order Bessel high-pass filter with gain 2.3, and at the end a second order Bessel low pass filter with unit gain, as shown in Fig. 3.

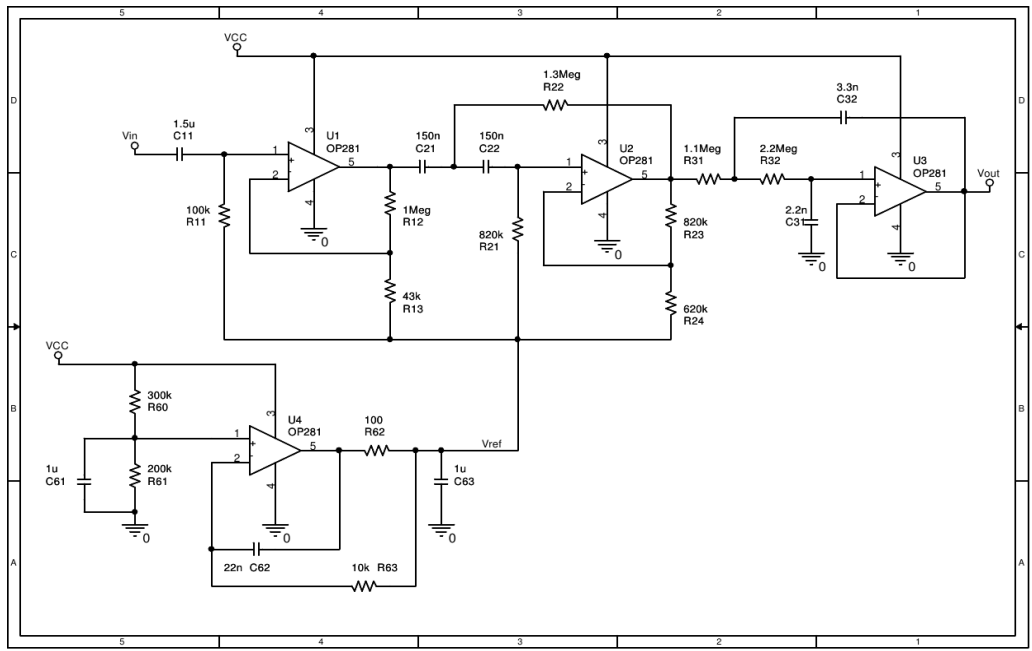

Figure 3. Band-pass active filter.

Both second-order Bessel filters have been realized using Sallen-Key circuital topology with positive supply only operational amplifier. The high pass section cuts at 1.4 Hz (-3 dB), reaching -40 dB at 0.22 Hz, while the low pass section cuts at 31 Hz (-3 dB) reaching -40 dB at 390 Hz. The high cut has been chosen to allow a good peak detection and suppression of any pickup noise.

The positive supply only circuit needs a voltage reference which is obtained using another operational amplifier. Filter circuit has been optimized for low power, using high value resistors and low power operational amplifier, obtaining power dissipation below 130 μW.

The start-up time is about 1 s, which is negligible if compared to the measurement time. The noise has been evaluated to be almost negligible at 12 bit acquisition.

All these parameters have been verified to be stable versus component value changes due to production tolerances. Most resistors have been requested at 5% and most capacitors have been requested at 20%. Statistical analysis of the circuit properties shows that, in case that tighter tolerance should be requested, it would be enough to use some 1 % tolerance resistors and 10 % capacitors.

Although the circuit is powered at $V_{cc}$ = 5 V, the reference level $V_{ref}$ has been put at 2V (a bit lower than $V_{cc}/2$) since output starts distorting at 4.2V and signal has positive peak larger than negative peaks. This signal level range comply with ADC input on ADuC812 microcontroller when the ADC reference is driven by $V_{cc}$ (5V), used as an external reference.

Since the pressure sensor output is an amplified signal coming from a full measurement bridge, its output is proportional to the supply voltage, hence correct pressure measurement requires ratiometric method. Moreover, since the ADC output values are the input signals divided by the reference voltage, ratiometric measurements are simply obtained using as ADC voltage reference the sensor supply voltage, which is also used for generating the voltage reference at the filter stages.

The frequency response of the designed pass-band filter is shown in Fig. 4.

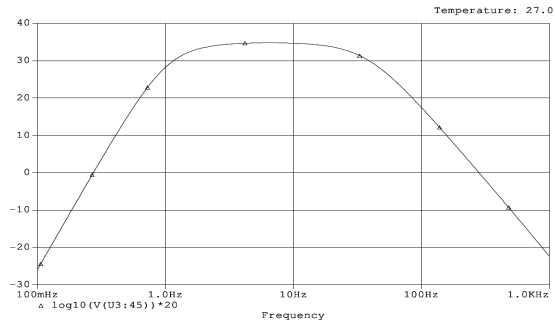

Figure 4. Frequency response of the active filter.

Fig. 5 shows the oscillometric signal, which is obtained by applying the narrow pass-band active filter to the pressure signal coming from the sensor-transducer system.

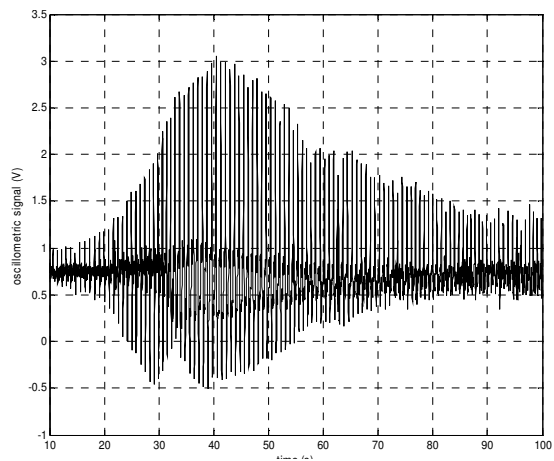

Figure 5. Oscillometric signal after the filter.

The microcontroller reads both the signal from the transducer and the output signal of the pass-band filter, thus evaluating the mean arterial pressure and the related values corresponding to systolic and diastolic blood pressure.

As already specified, we have used an Analog Devices AduC812 microcontroller, characterized by analog single-ended 8 acquisition channels, each of them multiplexed to a 12 bit ADC converter, flash, RAM, and UART serial interface. For pressure measurements, we use only two channels to process the input signals, which are sampled at different frequencies, buffered in the internal memory and analyzed on the fly. The microcontroller is charged with measurement scheduling, result memorizing and transmission using a GPRS modem. To this purpose, we have used a transmitting module employing a surface acoustic wave (SAW) transmitter to produce a carrier wave at the free frequency of 433.92 MHz at a voltage supply of 5 V and an absorbed current of 4 mA.

The chosen receiver is characterized by a high selectivity and insensitivity to electromagnetic fields, a working centre frequency of 433.92 MHz and a pass-band of 600 kHz. The typical values of the voltage supply and absorbed current are +5 V and 3 mA, respectively. Moreover, the receiver is characterized by a high sensitivity of -100 dBm which can be lowered in order to provide a suitable noise level reduction. Finally, we have tested our system by using a Lu-La Logic Analyzer, which has two input channels. We have connected the channel 0 to the receiver output and the channel 1 to the transmitter input, as shown in Fig. 6.

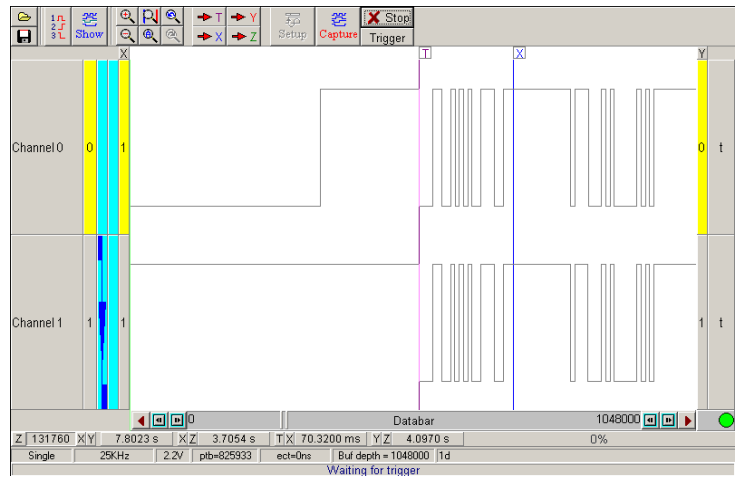

Figure 6. Testing stage by using a Lu-La Logic Analyzer. Channel 0 and 1 show the receiver output and transmitter input, respectively.

As shown in Fig. 6, the transmitted signal is perfectly recognized by the receiver.

## 4. CONCLUSIONS AND FUTURE DEVELOPMENTS

In this paper, a system to measure the arterial blood pressure is presented. Our device, based on the use of the oscillometric method, solves all the problems related to the typical approach for detecting the Korotkoff sounds. We have used a pressure sensor, an active filter and an ADC on board of the microcontroller ADuC812. Moreover, our device can perform the transmission of the measured pressure values to a remote computer using a GPRS modem.

Actually we are working to obtain a drastic reduction in dimensions of our prototype using industrial technologies.

## REFERENCES

[1]     Marani R., Perri A. G., (2011)  "Biomedical Electronic Systems to Improve the Healthcare Quality and Efficiency", In *Biomedical Engineering, Trends in Electronics, Communications and Software*, Ed. A. Laskovski, IN-TECH, ISBN 978-953-307-475-7, pp. 523 – 548, Available: http://www.intechweb.org

[2]     Marani R., Gelao G., Perri A.G., (2010) "High Quality Heart and Lung Auscultation System for Diagnostic Use on Remote Patients in Real Time", *The Open Biomedical Engineering Journal*, Vol. 4, pp.250-256.

[3]     Marani R., Perri A. G., (2010) "An Electronic Medical Device for Preventing and Improving the Assisted Ventilation of Intensive Care Unit Patients", *The Open Electrical & Electronic Engineering Journal*,  vol.4, pp.16-20.

[4]     Marani R., Perri A. G., (2010) "A new pressure sensor-based electronic medical device for the analysis of lung sounds", in *Proceedings of MELECON 2010*, Valletta, Malta.

[5]     Marani R., Gelao G., Perri A.G., (2010)  "A New System for Continuous Monitoring of Breathing and Kinetic Activity"; *Journal of Sensors*, Hindawi Publishing Corporation, vol. 2010, doi 10.1155/2010/434863/JS.

[6]     Gelao G.,  Marani R., De Leonardis F., Passaro V.M.N., Perri A. G., (2011)  "Architecture and Front-end for in-vivo blood glucose sensor based on impedance spectroscopy", in *Proceedings of IWASI 2011, 4$^{th}$ IEEE International Workshop on Advances in Sensors and Interfaces*, Savelletri di Fasano, Brindisi, Italy,  pp. 139-141.

[7]     Marani R., Gelao G., Perri A.G., (2012) "Design and Prototyping of a Miniaturized Sensor for Non-Invasive Monitoring of Oxygen Saturation in Blood", *International Journal of Advances in Engineering & Technology*, Vol.2, issue 1,  pp. 19-26.

[8]    Gelao G., Marani R., Carriero V., Perri A. G., (2012) "Design of a Dielectric Spectroscopy Sensor for Continuous and Non-Invasive Blood Glucose Monitoring", *International Journal of Advances in Engineering & Technology*, Vol. 3, issue 2, pp. 55-64.

[9]    Wagner S., Toftegaard T.S., Bertelsen O., (2012) "Challenges in Blood Pressure Self-Measurement", *International Journal of Telemedicine and Applications*, vol. 2012, Article ID 437350, doi:10.1155/2012/437350, pp.1-8.

[10]   Pickering T. G., (1996) "Recommendations for the use of home (self) and ambulatory blood pressure monitoring", *American Journal of Hypertension*, vol. 9, no. 1, pp. 1-11.

[11]   Kearney P.M., Whelton M., Reynolds K., Muntner P., Whelton P.K., He J., (2005) "Global burden of hypertension: analysis of worldwide data", *The Lancet*, vol. 365, no. 9455, pp. 217-223.

[12]   Hansen N. E., Haunsø S., Schaffalitzky de Muuckadell O.B., (2005) "Hypertensio arterialis", *Medicinsk kompendium*, Nyt Nordisk Forlag Arnold Busck, Copenhagen, Denmar, pp. 314-315.

[13]   Pickering T. G., Miller N. H., Ogedegbe G.,. Krakoff L. R, Artinian N.T., Goff D., (2008) "Call to action on use and reimbursement for home blood pressure monitoring: executive summary: a joint scientific statement from the American Heart Association, American Society of Hypertension, and Preventive Cardiovascular Nurses Association", *Hypertension*, vol. 52, no. 1, pp. 1-9.

[14]   Chobanian A. V.,. Bakris G.L, Black H.R., (2003) "Seventh report of the joint national committee on prevention, detection, evaluation, and treatment of high blood pressure", *Hypertension*, vol. 42, no. 6, pp. 1206-1252.

[15]   T. Tagawa T., T. Tamura T., Ake Öberg P., (1997) "Biomedical Transducers and Instruments", CRC.

[16]   Ball-llovera A., Del Rey R., Ruso R., Ramos J., Batista O., Niubo I., (2003) "An experience in Implementing the Oscillometric Algoritm for Non-Invasive Determination of Human Blood Pressure", in *Proceedings of 25th International Conference of IEEE*, Enginnering in medicine and biology society, Vol. 4, pp 3173-3175.

[17]   Marani R., Perri A. G., (2012)  "Design of Advanced Electronic Biomedical Systems"; *International Journal of Advances in Engineering & Technology*, Vol. 4, issue 1, pp. 15-25.

## Authors

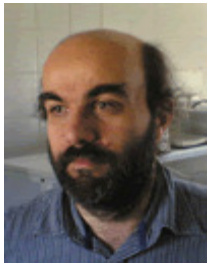

**Gennaro Gelao** received the Laurea degree in Physics from University of Bari, Italy, in 1993 and his Ph.D. degree in Physics in 1996 at CERN.
From 2001 he cooperates with the Electronic Device Laboratory of Polytechnic University of Bari as technical manager. Actually he works for the design, realization and testing of nanometrical electronic systems, quantum devices and CNTFETs.
Moreover his research activities also includes the design, modelling and experimental characterization of devices and systems for biomedical applications.
Dr. Gelao has published over 80 papers.

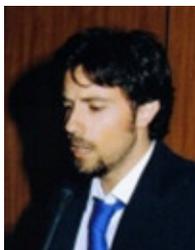

**Roberto Marani** received the Master of Science degree (*cum laude*) in Electronic Engineering in 2008 from Polytechnic University of Bari, where he received his Ph.D. degree in Electronic Engineering in 2012.
He worked in the Electronic Device Laboratory of Bari Polytechnic for the design, realization and testing of nanometrical electronic systems, quantum devices and FET on carbon nanotube. Moreover Dr. Marani worked in the field of design, modelling and experimental characterization of devices and systems for biomedical applications.
In December 2008 he received a research grant by Polytechnic University of Bari for his research activity. From February 2011 to October 2011 he went to Madrid, Spain, joining the Nanophotonics Group at Universidad Autónoma de Madrid, under the supervision of Prof. García-Vidal.
Currently he is involved in the development of novel numerical models to study the physical effects that occur in the interaction of electromagnetic waves with periodic nanostructures, both metal and dielectric. His research activities also include biosensing and photovoltaic applications.
Dr. Marani is a member of the COST Action MP0702 - Towards Functional Sub-Wavelength Photonic Structures, and is a member of the Consortium of University CNIT – Consorzio Nazionale Interuniversitario per le Telecomunicazioni. Dr. Marani has published over 100 scientific papers.

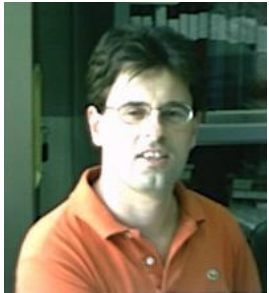

**Vittorio M. N. Passaro** received the Laurea degree (*cum laude*) in electronic engineering from the University of Bari, Italy, in 1988, and the Ph.D. degree in electronic engineering from Politecnico di Bari in 1992. He joined Politecnico di Bari as an Associate Professor of electronic technologies and photonics since October 2000. He is involved in modeling, design and simulation of a number of integrated optical devices and circuits for optical signal processing, telecommunications and sensing, by considering a number of materials, in particular ferroelectric and semiconductors (silicon and compounds, III–V alloy compounds). Currently, he is mainly involved in silicon photonics. Since 2004, he formed and led the Photonics Research Group at Politecnico di Bari. He is the Editor of five international books, as well as the author or co-author of more than 290 papers published in international journals and conference proceedings, with more than 1700 cites in scientific literature. He holds two international patents. His current research interests include several theoretical and experimental aspects of optoelectronic and photonic technologies. Since 2012 he is the Editor-in-Chief of Sensors journal.

Prof. Passaro is Senior Member of the Optical Society of America since 2012, IEEE Senior Member since 2005, and Associate Member of National University Consortium for Telecommunications (CNIT) and National Institute for Nuclear Physics (INFN).

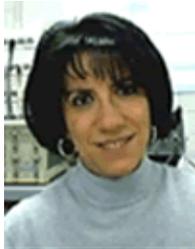

**Anna Gina Perri** received the Laurea degree *cum laude* in Electrical Engineering from the University of Bari in 1977. In the same year she joined the Electrical and Electronic Department, Polytechnic University of Bari, Italy, where she is Full Professor of Electronics from 2002.

From 2003 she has been associated with the National Institute of Nuclear Phisics (INFN) of Napoli (Italy), being a part of the TEGAF project: "Teorie Esotiche per Guidare ed Accelerare Fasci", dealing with the optimal design of resonance-accelerating cavities having very high potentials for cancer hadrontherapy.

In 2004 she was awarded the "Attestato di Merito" by ASSIPE (ASSociazione Italiana per la Progettazione Elettronica), Milano, BIAS'04, for her studies on electronic systems for domiciliary teleassistance.

Her current research activities are in the area of numerical modelling and performance simulation techniques of electronic devices for the design of GaAs Integrated Circuits and in the characterization and design of optoelectronic devices on PBG (Phothonic BandGap).

Moreover she works in the design, realization and testing of nanometrical electronic systems, quantum devices, FET on carbon nanotube and in the field of experimental characterization of electronic systems for biomedical applications.

Prof. Perri is the Head of the Electron Devices Laboratory of the Polytechnic University of Bari.

She has been listed in the following volumes: Who's Who in the World and Who's Who in Engineering, published by Marquis Publ. (U.S.A.).

She is author of over 250 journal articles, conference presentations, twelve books and currently serves as a Referee of a number of international journals.

Prof. Perri is the holder of two italian patents and the Editor of two international books.

She is also responsible for research projects, sponsored by the Italian Government.

Prof. Perri is a member of the Italian Circuits, Components and Electronic Technologies – Microelectronics Association, and an Associate Member of National University Consortium for Telecommunications (CNIT).

Prof. Perri is a Member of Advisory Editorial Board of International Journal of Advances in Engineering & Technology (IJAET) and of Current Nanoscience (Bentham Science Publishers).